\documentclass[twocolumn,aps,prd,nofootinbib,tightenlines]{revtex4}

\usepackage{amsmath, amsfonts, amsthm, amssymb, graphicx}

\allowdisplaybreaks[3]

\def\be{\begin{equation}}
\def\ee{\end{equation}}
\def\ba{\begin{eqnarray}}
\def\ea{\end{eqnarray}}
\newcommand{\nn}{\nonumber\\}
\newcommand{\ud}{\mathrm{d}}
\def \pd {\partial}

\begin{document}

\title{Note on the Stabilities of the Light-like Galileon Solutions}

\author{Shuang-Yong Zhou}
\email[]{ppxsyz@nottingham.ac.uk}

\affiliation{School of Physics and Astronomy,
University of Nottingham, Nottingham NG7 2RD, UK}

\date{\today}

\begin{abstract}

Light-like galileon solutions have been used to investigate the chronology problem in galileon-like theories, and in some cases may also be considered as solitons, evading a non-existence constraint from a zero-mode argument. Their stabilities have been analyzed via ``local'' approximation, which appears to suggest that all these light-like solutions are stable. We re-analyze the stability problem by solving the linear perturbation equation \emph{exactly}, and point out that the finite energy condition is essential for the light-like solitons to be stable. We also clarify potential ghost instabilities and why the zero-mode argument can not be naively generalized to include the light-like solitons.

\end{abstract}

\maketitle

\section{Introduction}

The galileon model is a scalar field theory that possesses a novel field symmetry $\pi \to \pi + b_\mu x^\mu + c$, $b_\mu, c = {\rm const}$ \cite{Nicolis:2008in}. It is originally constructed to describe large distance modifications of General Relativity \cite{Nicolis:2008in}, which can capture some local 4D gravitational physics of certain braneworld scenarios \cite{Luty:2003vm, Nicolis:2004qq, deRham:2010eu, Dvali:2000hr, Deffayet:2009wt} as well as the non-linear massive gravity model \cite{deRham:2010kj} (see \cite{Clifton:2011jh, Trodden:2011xh, Qiu:2011cy} and references therein for details and related cosmology). Albeit simple, this novel field theory had been rarely explored before, but has been since found to have interesting field-theoretic properties. The general galileon Lagrangian in $(n+1)$-D is given by
\be
\mathcal{L} = -\pd_{\mu} \pi \pd^\mu \pi
- \sum_{i=2}^{n+1} \alpha_i \pd_{\mu_1} \pi \pd^{[\mu_1} \pi
 \pd_{\mu_2} \pd^{\mu_2} \pi \cdots \pd_{\mu_i} \pd^{\mu_i]} \pi  ,
\ee
where $[~]$ means anti-symmetrization of the indices and $\alpha_i$'s are the couplings of the theory. Despite the apparent high order derivatives, its equation of motion remains second order, avoiding the Ostrogradski ghosts. Also, perhaps more interestingly, all the couplings in this model are not renormalized by loop corrections \cite{Nicolis:2004qq, Hinterbichler:2010xn}. So it is interesting to investigate its classical non-linear dynamics, particularly to search for stable solitonic solutions. As the non-linear galileon terms contain higher order gradients, it may be possible that Derrick's theorem \cite{Derrick:1964ww} can be evaded. However, using a zero-mode argument, Endlich {\it et al} \cite{Endlich:2010zj} have shown that there is no stable static soliton in the single galileon model. This may be evaded by a multi-field generalization coupled with a non-linear sigma model constraint \cite{Padilla:2010ir, Deffayet:2010zh}, along the lines of the skyrmions \cite{Skyrme:1961vq, Skyrme:1962vh}, which have been used to model baryons and resonance states.

Another way to evade the zero-mode constraint, as recently pointed out by Masoumi and Xiao \cite{Masoumi:2012vz}, is to construct solutions that can not be boosted to a rest frame, i.e., a solution that is light-like or wave-like (see also \cite{Evslin:2011vh, Burrage:2011cr, Evslin:2011rj}). In 1+1D\;\footnote{In this note, we will mainly focus on 1+1D for simplicity. In 3+1D one may equivalently consider homogeneous 2-brane or domain wall configurations.}, the only non-linear galileon term is the $\alpha_2$ coupling, in which case it has been found that \emph{any} left or right moving mode $\pi=\pi_0(x\pm t)$ is a solution to the equation of motion  \cite{Masoumi:2012vz, Evslin:2011vh, Burrage:2011cr, Evslin:2011rj}. The reason for this can be understood via the ``target space diffeomorphism'' symmetry \cite{Fairlie:1992yy, Fairlie:2011md}. Or, one may simply see it from the equation of motion in the light cone coordinates $(u=x+t,v=x-t)$,
\be \label{eombg}
\pd_u\pd_v \pi + 2 \alpha \left[ (\pd_u \pd_v \pi)^2 - \pd_u^2 \pi \pd_v^2 \pi \right] = 0   ,
\ee
where $\alpha\equiv 3\alpha_2/4$. The null modes $\pi=\pi_0(x\pm t)$ vanish the linear and non-linear pieces simultaneously. If the solution (the field derivative, not necessary the field itself, as will be clear later) is a concentrated wavepacket, we may consider it as a soliton. This is in contrast to a typical soliton where the non-linear terms cancel the linear terms in the equation of motion. Therefore, galileon solitons can form even when the non-linear interactions are small.

These light-like solutions have been used to investigate closed time-like curves and their implications for the chronology protection problem in various theories \cite{Evslin:2011vh, Burrage:2011cr, Evslin:2011rj}. However, so far, the stabilities of these light-like solutions have only been tackled in the ``local'' approximation, i.e., by calculating the eigenvalues of the kinetic coefficient matrix of the perturbative Lagrangian, based on which the light-like solutions appear to be stable for \emph{any} light-like configurations $\pi=\pi_0(x\pm t)$ \cite{Masoumi:2012vz, Evslin:2011rj}, putting them on the same footing as the linear wave equation. As one may expect, this ``local'' stability analysis is by no means rigorous (see e.g.~\cite{Nicolis:2004qq, Endlich:2011vg}), and this peculiar conclusion may be merely an artifact of this approximation. This, however, can only be determined if the linear perturbation equation in question can be solved exactly, which typically is difficult. In this note, we will re-examine the stability problem of the light-like galileon solutions, and show that the relevant linear perturbation equation is fortunately \emph{solvable}. The finite energy condition of the given solution, which has been largely neglected so far, is important for the solution to be stable. There may be tachyonic instabilities (or growing modes) if the light-like configurations are not properly chosen to have finite energies. Also, we will point out that ghost instabilities will develop for certain configurations, which has been neglected in \cite{Masoumi:2012vz}, and clarify why the zero-mode argument of \cite{Endlich:2010zj} can not be naively generalized to include the light-like soliton case.

\section{Linear Stability of the light-like Solutions}

As discussed in the Introduction, any light-like configuration $\pi_0(x \pm t)$ is a solution to the galileon equation of motion (\ref{eombg}). Since we are only interested in the linear stability, we perturb a given solution $\pi=\pi_0(x\pm t)+\phi$ and obtain the perturbative action
\be   \label{actp}
\delta_2 S  = \int \ud^2 x\; Z^{\mu\nu} \pd_\mu \phi \pd_\nu \phi   .
\ee
The kinetic coefficients are
\begin{align}
Z^{\mu\nu} &= -\eta^{\mu\nu} + 2\alpha \pd^\mu \pd^\nu \pi_0
\nn
        &  =
\left(
 \begin{array}{cc}
   1+a(x\pm t) & \mp a(x\pm t) \\
       \mp a(x\pm t) & -1+a(x\pm t) \\
\end{array}
\right)   ,
\end{align}
where we have defined the function $a(x\pm t)=2\alpha\ddot{\pi_0}(x\pm t)$. From the perturbative action (\ref{actp}), we get the linear perturbative equation of motion:
\be \label{keyprob}
Z^{\mu\nu} \pd_\mu \pd_\nu \phi = 0  .
\ee
where we have used the relation $\pd_\mu Z^{\mu\nu} = 0$, thanks to the galileon construction. So the problem is to solve Eq.~(\ref{keyprob}) exactly. Note that Eq.~(\ref{keyprob}) is of the hyperbolic type, since the discriminant $\Delta=Z^{00} Z^{11}-(Z^{01})^2=-1<0$, so it has a well-defined Cauchy initial problem.

\subsection{Tachyonic instabilities} \label{tachins}

First, we will be looking at the tachyonic instabilities, that is, whether there are modes that grows (quasi-)exponentially with time. Unable to solve the second order partial differential equation exactly, this is often tackled in the ``local'' approximation, i.e., by assuming the background varies slowly both spatially and temporally so that the kinetic coefficients $Z^{\mu\nu}$ can be treated as constants. This is justified if we are only concerned about the ultraviolet modes: When we ``zoom in'' at a given spacetime point, the smooth functions $Z^{\mu\nu}(x^\mu)$ certainly vary slowly. Then the linear stability problem is reduced to determining whether  $Z^{\mu\nu}$ has the right Lorentz signature at every spacetime point, or, equivalently  whether $Z^{0i}Z^{0j}-Z^{00}Z^{ij}$ is positive definite \cite{Nicolis:2004qq}. For the problem at hand, $Z^{\mu\nu}$ indeed has the right signature, independent of the form of $a(x\pm t)=2\alpha \ddot{\pi_0}(x\pm t)$, as observed by \cite{Masoumi:2012vz, Endlich:2011vg}. However, this ``local'' approximation may miss some important instabilities such as resonances with the background. For the problem at hand, as one may expect, the stability problem being independent of $a(x\pm t)$ could well only be an artifact of this approximation. So solving the linear perturbation problem \emph{exactly}, if possible, is certainly desirable, and  fortunately, is achievable for the problem in question.

We will solve it in the light cone coordinates ($u=x+t$, $v=x-t$), where the perturbative equation of motion can be written as,
\begin{align}
a(u) \pd_v^2 \phi -\pd_u \pd_v \phi &= 0  \qquad{\rm for}\quad \pi_0 = \pi_0(x+t),
\\
\label{uvrmeq}
a(v) \pd_u^2 \phi -\pd_u \pd_v \phi  &= 0  \qquad{\rm for}\quad \pi_0 = \pi_0(x-t),
\end{align}
respectively for the left and right moving modes. As expected, these equations are symmetric in terms of exchanging $u$ and $v$.

Without loss of generality, we can focus on the right moving mode $\pi_0=\pi_0(x-t)$, that is, Eq.~(\ref{uvrmeq}). To solve Eq.~(\ref{uvrmeq}), it is convenient to extend $\phi$ to be complex valued and separate the variables $\phi(u,v)=U(u)V(v)$. This leads to
\be
\frac{\pd_u^2 U(u)}{\pd_u U(u)} = \frac{\pd_v V(v)}{a(v)V(v)} = \lambda   ,
\ee
where $\lambda$ is a constant, or an eigenvalue for the relevant eigenmode. Note that here we should allow $\lambda$ to sample the whole complex plane. Now, we can solve $U(u)$ and $V(v)$ separately, whose general solutions are respectively given by
\begin{align}
U(u) &= c_1 +c_2 e^{\lambda u}   ,
\\
V(v) &= c_3 e^{\lambda b(v)}   ,
\end{align}
where $c_1, c_2, c_3$ are integration constants and the function
\be
b(v) \equiv \int^v a(v') \ud v' = 2 \alpha \pd_v \pi_0(v) .
\ee
Thus, in the $(t,x)$ coordinates, the eigenmodes of $\phi$ are
\be \label{eigenModes}
\phi^{(1)}_{\lambda_1}(t,x) = e^{\lambda_1 b(x-t)}, \quad
\phi^{(2)}_{\lambda_2}(t,x) = e^{2\lambda_2 t}e^{\lambda_2 B(x-t)} ,
\ee
where
\be
B(x-t) \equiv b(x-t)+x-t.
\ee
At this stage, $\lambda_1$ and $\lambda_2$ are arbitrary complex numbers and the most general solution to Eq.~(\ref{uvrmeq}) is the linear combination of all the these eigenmodes with different $\lambda_1$ and $\lambda_2$, which can be written as
\be \label{solspace}
\phi(t,x) = \sum_{ \lambda}  \left[ \rho_1(\lambda) \phi^{(1)}_{\lambda}(t,x)
+  \rho_2(\lambda) \phi^{(2)}_{\lambda}(t,x) \right] ,
\ee
where $\rho_1$ and $\rho_2$ are distribution functions of the eigenmodes, yet to be constrained by the initial and boundary conditions of the given problem. Only eigenmodes that can pass the ``filtering'' of the initial and boundary conditions can be used to construct the solution space of the given problem. If there are (quasi-)exponentially growing modes that survive the ``filtering'', then the background solution in question is unstable. Otherwise, it is free of tachyonic instabilities.

The initial and boundary conditions in question are that the perturbation field $\phi(t,x)$ and its ``velocity'' $\dot{\phi}(t,x)$ should be small, or bounded, initially. Since neither the eigenmodes of $\phi(t,x)$ nor those of $\dot{\phi}(t,x)$ degenerate at $t=0$, this requires that every eigenmode of $\phi^{(i)}_{\lambda_i}(t,x)$ and its ``velocity'' $\dot{\phi}^{(i)}_{\lambda_i}(t,x)$ are bounded at $t=0$. (Note that degeneration of eigenmodes at $t=0$ happens, for example, for the case of the 1+1D wave equation, in which case the reduction to the eigenmodes is not quite appropriate. We will discuss this in the Appendix.). Further, since $\phi^{(i)}_{\lambda_i}(t=0,x)$ and $\dot{\phi}^{(i)}_{\lambda_i}(t=0,x)$ are smooth functions, it is enough to require $\phi^{(i)}_{\lambda_i}(t=0,x\to \pm \infty)$ and $\dot{\phi}^{(i)}_{\lambda_i}(t=0,x\to \pm \infty)$ to be bounded, and these are our ``filters''. Any eigenmode, either itself or its ``velocity'', that is not bounded initially should be thrown away. The eigenmodes and their ``velocities'' at time $t=0$ are
\begin{align}
\phi^{(1)}_{\lambda_1}(0,x) = e^{\lambda_1 b(x)},  &\quad  \phi^{(2)}_{\lambda_2}(0,x) = e^{\lambda_2 [B(x)]} ,\\
\label{secondFilterT0}
\dot{\phi}^{(1)}_{\lambda_1}(0,x) \propto a(x) e^{\lambda_1 b(x)},  &~  \dot{\phi}^{(2)}_{\lambda_2}(0,x) \propto(a(x)-1) e^{\lambda_2 [B(x)]}  ,
\end{align}
where, as before, $a(x)=b'(x)$ and $B(x)=b(x)+x$. So whether these eigenmodes can pass through the ``filters'', consequently the stability of the problem, depends on the form of the background configuration $b(x)$, as well as the values of $\lambda_1$ and $\lambda_2$.

Note that if there are eigenmodes with pure imaginary $\lambda_1$ and $\lambda_2$ passing though the ``filters'', then the perturbations will oscillate around the background solution, i.e., the given solution will slightly ``breath in and out'' once perturbed. This is like a ``limit cycle'' in a standard dynamical system, which is regarded as stable. Potential instabilities arise only when $\lambda_1$ and $\lambda_2$ with non-zero real parts can pass though the ``filters''.

Let us first focus on the \emph{first ``filter''}: the eigenmodes $\phi^{(i)}_{\lambda_i}(t,x)$ should be bounded at $(t=0,x\to \pm \infty)$.

First, we note that whenever $\phi^{(1)}_{\lambda_1}(0,x) = e^{\lambda_1 b(x)}$ is bounded, $\phi^{(1)}_{\lambda_1}(t,x) = e^{\lambda_1 b(x-t)}$ is also bounded, so there are no growing modes from the $\phi^{(1)}_{\lambda_1}(t,x)$ sector.

The $\phi^{(2)}_{\lambda_2}(t,x)$ sector is more complicated. Since the first ``filter'' requires $e^{\lambda_2 B(x)}$, hence also $e^{\lambda_2 B(x-t)}$, to be bounded, we see from Eq.~(\ref{eigenModes}) that modes with $Re(\lambda_2)>0$ are (quasi-)exponentially growing. (The potential loophole for this argument is that, even for $Re(\lambda_2)>0$, when $t\to +\infty$, it is possible that $e^{\lambda_2 B(x-t)} \to 0$ at a rate such that $e^{2\lambda_2 t}e^{\lambda_2 B(x-t)}$ is bounded. However, this is only true for certain regions of $x$. Indeed, as $B(x-t)$ is a regular function of $x-t$, there are always regions of $x$ such that $B(x-t)$, hence also $e^{\lambda_2 B(x-t)}$, remains finite.) If only modes with $Re(\lambda_2)\leq 0$ can pass through the first ``filter'', without the need of examining the second ``filter'', we can say that there are no tachyonic modes for the given configuration $b(x-t)$. To proceed, let us classify the initial configuration $B(x)$ according to different asymptotes:
\begin{enumerate}

\item \emph{$B(x\to \pm \infty)$ approaches $-\infty$ but not $+\infty$:} modes with $Re(\lambda_2)\geq 0$ can pass though, thus this case is potentially \emph{unstable}.

\item \emph{$B(x\to \pm \infty)$ approaches $+\infty$ but not $-\infty$:} only modes with $Re(\lambda_2)\leq 0$ can pass though, thus this case is \emph{stable}, without the need of examining the second ``filter''.

\item \emph{$B(x\to \pm \infty)$ approaches both $-\infty$ and $+\infty$:} only modes with $Re(\lambda_2) = 0$ can pass though, thus this case is \emph{stable}, without the need of examining the second ``filter''.

\item \emph{$B(x\to \pm \infty)$ are bounded:} modes with any $\lambda_2$ can pass though, thus this case is potentially \emph{unstable}.

\end{enumerate}

Now, let us consider the \emph{second ``filter''}: $\dot{\phi}^{(i)}_{\lambda_i}(t,x)$ should be bounded at $(t=0,x\to \pm \infty)$. From Eq.~(\ref{secondFilterT0}), we can see that this ``filter'' is equivalent to requiring that
\be \label{filter2}
\pd_x\left[\phi^{(1)}_{\lambda_1}(0,x=\pm \infty)\right]
~~{\rm and}~~
\pd_x\left[\phi^{(2)}_{\lambda_2}(0,x=\pm \infty)\right]
\ee
are bounded. Note that this is \emph{not} implied by the requirement that $\phi^{(1)}_{\lambda_1}(0,x=\pm \infty)$ and $\phi^{(2)}_{\lambda_2}(0,x=\pm \infty)$ are bounded, as one would have naively thought. For example, while $\sin(x^2)$ is bounded within $(-\infty,+\infty)$, its derivative is not. Further classification of $B(x)$ is rather complicated, whose details we shall not go into. We just point out that, depending on the background solution $b(x)$, the second ``filter'' may kill off more unstable modes, as well as stable modes, which have passed through the first ``filter''. Once the form of the background solution is given, the condition (\ref{filter2}) can be readily checked and the full stability problem is therefore easily determined.

However, it is not difficult to find examples where unstable modes survive after passing through both of the ``filters''. For example, we can choose $b(x)=c(x)-x$ where $c(x)$ is bounded, for which case $\lambda_1=0$ and arbitrary $\lambda_2$ can pass through the ``filters''.

So we see that not all light-like (or wave-like) solutions $\pi=\pi_0(x-t)$ are stable, in contrast to what one would infer from the analysis of the ``local'' approximation.

So far, we have been considering general solutions, solutions that may or may not have non-trivial global structures. This is important if one is interested in, for example, globally oscillatory solutions or processes involving extreme long wavelengths. However, solitons are normally defined as solutions that are localized and have finite energies. If we are interested in localized solutions with finite energies, then we will see that galileon solitons in 1+1D are always stable. Note that it is not necessarily for the galileon field itself to be localized, but its derivative has to fall off quickly enough so that the energy
\be
E[\pi_0]=2\int \ud v (\pd_v \pi_0(v))^2 = \frac{1}{2\alpha^2}\int \ud v\; b(v)^2
\ee
is finite. This requires $b(v)$ to fall off quickly at infinities, so we have $B(x\to \pm \infty)=x$. Therefore, the finite energy soliton case corresponds to the case 3 above,  which is fortunately stable. From the first ``filter', we know that $\lambda_2$ has to be pure imaginary in this case. Also, since $b(v)$ has to fall off quickly at infinities, the second ``filter'' is automatically satisfied. So, if perturbed, galileon solitons will ``breath in and out'' during their evolutions.

\subsection{Ghost instabilities}

The kinetic term of the action (\ref{actp}) is $Z^{00}\dot{\phi}^2 = (1+a(v))\dot{\phi}^2$ (Again, without lost of generality, we have focused on the right moving mode.). Clearly, when $a(v)<-1$, there is a ghost instability, which was neglected in  \cite{Masoumi:2012vz}. To avoid the ghost instability, we require
\be \label{cond1}
a(v) = 2\alpha \ddot{\pi_0}(v)  \geq -1.
\ee
This does not contradict the fact that $Z^{\mu\nu}$ is negative definite, because the signatures of the temporal and spatial components can swap. If the scalar $\pi$ is a standing only field, a ghost instability is not really a problem {\it per se}. After all, the action can always be re-defined by multiplying an overall sign, which does not change the equation of motion and leads to a well-defined path integral formulation. But when $\pi$ couples to other fields such as gravity, a ghost signals very serious instabilities, particularly in the quantum theory (see e.g.~\cite{Cline:2003gs}).

\section{Evasion of the Zero-mode Argument}

In \cite{Masoumi:2012vz}, it was pointed out that the zero-mode argument of \cite{Endlich:2010zj} does not include the case where the soliton can not be brought to a rest frame. But one might wonder whether we can apply a similar instability argument to the light-like solitons. After all, a soliton moving at the speed of light can still be invariantly translated spatially or temporally. Here we explicitly show why similar zero mode arguments can not be naively applied to the light-like solitons.

First, we review Endlich {\it et al}'s zero-mode argument for  {\it static} solitons in the single galileon theory  \cite{Endlich:2010zj}. Suppose there exists a solitonic solution which can be boosted to be static $\pi=\pi_0(x^i)$ (In this section we work on general $n+1$D dimensions.). To decide whether this soliton is stable or not, one perturbs the soliton $\pi=\pi_0(x^i)+\phi$ and see whether the energy of the soliton can be lowered by some perturbation $\phi$. For the galileon theory, the perturbative energy of the static soliton is given by
\be
\delta_2 H = \int \ud^n x ( \dot{\phi}^2 + Z^{ij} \pd_i \phi \pd_j \phi  )  ,
\ee
where
\be
Z^{ij} =  \delta^{ij} - 2\alpha (\pd_k \pd^k \pi_0 \delta^{ij}- \pd^i \pd^j \pi_0) + \cdots .
\ee
Since the energy of the soliton is invariant under spatial translations, the perturbation
\be
\phi = \pi_0(x^i+\epsilon^i)-\pi_0(x^i) = \epsilon^i\pd_i\pi_0 +\mathcal{O}(\epsilon^2)
\ee
should give rise to $\delta_2 H = 0 + \mathcal{O}(\epsilon^3)$, i.e., this is a zero energy mode. Now, we can use this zero mode to probe the positivity of the kinetic coefficients $Z^{ij}$. Far away from the center of the soliton, the definition of solitons requires $\pd\pi_0$, $\pd\pd\pi_0$ to be very small, so we have $Z^{ij}\simeq  \delta^{ij}$. This implies that $\delta_2 H$ gets positive $\mathcal{O}(\epsilon^2)$ contribution from the integration far away from the center of the soliton. For the zero energy mode, there must be some negative contribution near the center of the soliton, which implies that $Z^{ij}$ has negative eigenvalues near the center. The perturbation $\phi$, hence $\pd_i\phi$, is essentially initial and boundary conditions of the system that can be arbitrary chosen. If $\pd_i\phi$ are chosen to be deformations of the zero mode that ``enhance'' the negative eigenvalue directions of $Z^{ij}$, we can get negative $\delta_2 H$. Therefore, the assumed soliton is not stable\;\footnote{In $1+1$D, there is actually no non-trivial static solutions, as the equation of motion reduces to a trivial linear equation.}.

Now, why can we not develop a similar argument for the null solitons, using the spacetime translation symmetry the system still possesses? For the static soliton case, we were actually a bit sloppy in differentiating the perturbative Hamiltonian (Hamiltonian derived from the quadratic perturbative Lagrangian) and the full Hamiltonian (Hamiltonian derived from the full Lagrangian and then perturbed to quadratic order). The spacetime translation invariance is a symmetry of Minkowski space, i.e., a symmetry of the full Lagrangian or Hamiltonian, but not necessarily a symmetry for the perturbative Lagrangian or Hamiltonian. So the full Hamiltonian is the one we should really be using, if we want to make use of the spacetime translation symmetry. On the other hand, the zero-mode argument above relies on the use of the quadratic Hamiltonian. This is well justified as long as the background solution, which minimizes the Lagrangian, also minimizes the Hamiltonian, in which case the quadratic perturbative Hamiltonian and the full Hamiltonian only differ by a constant, the energy of the background solution. Whilst this is the case for a static background, it is not necessarily true for a time-dependent one such as a light-like soliton. Therefore, for a time-dependent background, apart from the zero-order piece and the quadratic piece, the full Hamiltonian may include a linear piece, and the zero-mode associated with the space translation does not necessarily vanish the quadratic Hamiltonian (up to second order in perturbation). Instead, it should vanish the linear and quadratic pieces combined. Therefore, the zero-mode argument generally does not apply to a time-dependent background.

We can check that this is indeed the case for our situation at hand. 
The full Hamiltonian of the 1+1D galileon theory is given by
\be
H =\int \ud x [(1+2\alpha \pi'')\dot{\pi}^2+\pi'^2 ]   .
\ee
Perturbing it ($\pi=\pi_0(x-t)+\phi(t,x)$) up to quadratic order and integrating by parts, we have
\be
H   = H_0 +H_1+H_2
\ee
with
\begin{align}
    H_0 &=\int \ud x 2\dot{\pi_0}^2  , \nn
    H_1 &= \int \ud x [2(1+2\alpha\ddot{\pi_0})\dot{\pi_0} \dot{\phi}
                  -2(1-2\alpha\ddot{\pi_0})\dot{\pi_0} \phi' ]   , \nn
    H_2 &=\int \ud x [(1+2\alpha \ddot{\pi_0})\dot{\phi}^2+\phi'^2+4\alpha \dot{\pi_0}\dot{\phi} \phi'' ]   . \nonumber
\end{align}
Substituting in the zero-mode $\phi = \pi_0(x-t+\epsilon)-\pi_0(x-t) = -\epsilon \dot{\pi_0} + \epsilon^2\ddot{\pi_0}/2+\mathcal{O}(\epsilon^3)$, integrating by parts and collecting the $\mathcal{O}(\epsilon^2)$ terms, we have $H_1+\mathcal{O}(\epsilon^3)=-2\epsilon^2\int \ud x \ddot{\pi_0}^2 =-H_2+\mathcal{O}(\epsilon^3)$.

\section{Discussion}

In this note we have re-examined the (linear) stability problem of the light-like galileon solutions. These solutions may be considered as solitons in some cases and are interesting in their own right. They have also been used to investigate closed time-like curves and their implications for the chronology protection problem in various galileon-like modified gravity theories \cite{Masoumi:2012vz, Evslin:2011vh, Burrage:2011cr, Evslin:2011rj}. By solving the second order linear perturbation equation explicitly, we have been able to determine whether there are (quasi-)exponentially growing modes for given background configurations and classify general configurations according to their behaviors at the initial infinite boundaries. For the case of a finite energy soliton (or wavepacket), it is the finite energy condition of the soliton that guarantees the stability of the soliton, a point that has been overlooked so far. For a general solution with non-trivial global structures, the stability of the solution is more subtle. But if the configuration is given, its stability can be readily checked using the conditions established. Examining the positivity of the kinetic coefficient matrix is frequently used to determine the absence of tachyonic instabilities in linear perturbation theory. This mathematically solvable case exemplifies the expected limitation of this approach. We have also pointed out that for certain solitonic configurations ghost instabilities can also develop, which will become a serious problem when the galileon is coupled to other fields, and shown why the zero-mode argument of Endlich {\it et al} \cite{Endlich:2010zj} can not be naively generalized to include the light-like solitons.

Finally, we pass by a couple of curious observations about the galileon solitons. In this simple non-linear theory, we have a situation that localized, or wavepacket, solutions are stable while very long wavelength solutions are often susceptive to disintegration. If a system starts with a long wavelength dominated configuration, small wavepackets, or ``massless particles'', can emerge from it during its evolution, which may have interesting late time chaotic phenomena. This is in contrast to the linear wave equation, in which case even globally non-trivial solutions are stable. Also, the theory with only the $\alpha_2$ coupling is exactly the decoupling limit of the braneworld DGP model \cite{Luty:2003vm, Nicolis:2004qq}, where the extrinsic curvature of the DGP brane is roughly $K_{\mu\nu}\simeq \pd_\mu\pd_\nu \pi$. For the light-like solution, we have $K=\eta^{\mu\nu}K_{\mu\nu}\simeq \Box \pi_0(x\pm t)=0$, whose geometric origin may be interesting to investigate.

\appendix

\section{Stabilities of the 1+1D Wave Equation}

Any function $\psi_0(x\pm t)$ is a solution to the 1+1D wave equation $\ddot{\psi}-\psi''=0$. The perturbation on this solution $\varphi=\psi-\psi_0$ is also governed by the wave equation
\be
\ddot{\varphi}-\varphi''=0  .
\ee
Its eigenmodes are $\varphi^{(1)}_{\lambda_1}(t,x)=e^{\lambda_1(x+t)}$ and $\varphi^{(2)}_{\lambda_2}(t,x)=e^{\lambda_2(x-t)}$, so the general solution is given by
\be \label{waveEqGeneralSol}
\varphi(t,x)=\sum_{ \lambda} \left[ f_1(\lambda)\varphi^{(1)}_{\lambda}
+f_2(\lambda)\varphi^{(2)}_{\lambda} \right] .
\ee
To decide its stability, we can let this general solution to pass through the ``filtering'' of the initial and boundary conditions.

We first focus on the condition that $\varphi(t=0,x=\pm \infty)$ is bounded. Now, the eigenmodes $\varphi^{(1)}_{\lambda}(t,x)$ and $\varphi^{(2)}_{\lambda}(t,x)$ degenerate at $t=0$. In this case, we may not simply reduce the initial and boundary conditions for Eq.~(\ref{waveEqGeneralSol}) to a bunch of initial and boundary conditions for the eigenmodes individually. To see this, we do the ``filtering'' in the two ways and check whether they are consistent. First, impose the initial and boundary conditions that the eigenmodes $\varphi^{(1)}_{\lambda}(0,x=\pm \infty)=\varphi^{(2)}_{\lambda}(0,x=\pm \infty)$ are bounded. This essentially restricts $\lambda$ to be pure imaginary, and therefore it appears that all unstable eigenmodes have been ``filtered'' away. On the other hand, if we only require that $\varphi(t=0,x=\pm \infty)$ are bounded, there are unstable eigenmodes that can survive this ``filtering''. This can be achieved by designing special mode distributions. For example, if $f_1(\lambda)=-f_2(\lambda)$ is satisfied within $(-c,+c)$, $c$ being a positive real number, then modes with $\lambda\in (-c,+c)$ can also survive the ``filtering'' of $\varphi(t=0,x=\pm \infty)$ being bounded. But, from Eq.~(\ref{waveEqGeneralSol}), these modes are clearly unstable in terms of time evolution. Therefore, we see the two ways of ``filtering'' are inequivalent.

Nonetheless, any solution $\psi_0(x\pm t)$ to the wave equation is (by accident) actually stable, which can be seen by taking into account the second ``filtering'' condition that the field's initial ``velocity'' is also bounded.

{\bf Acknowledgement}

We thank Paul Saffin, Antonio Padilla and Clare Burrage for helpful discussions and the University of Nottingham for support.

\end{document}